\documentclass[%
 aip,
 amsmath,amssymb,
 reprint,%
]{revtex4-2}

\usepackage{graphicx}
\usepackage{dcolumn}
\usepackage{bm}
\usepackage{hyperref} 
\usepackage{epstopdf}
\newcommand{\ket} {\rangle}

\begin{document}

\preprint{APS/123-QED}

\title{Continuous Real-Time Detection of Quasiparticle Trapping in Aluminum Nanobridge Josephson Junctions}

\author{J. T. Farmer}
\affiliation{Department of Physics \& Astronomy, Dornsife College of Letters, Arts, \& Sciences, University of Southern California, Los Angeles, CA 90089, USA}
\affiliation{Center for Quantum Information Science \& Technology, University of
Southern California, Los Angeles, CA 90089, USA}
\author{A. Zarassi}
\affiliation{Department of Physics \& Astronomy, Dornsife College of Letters, Arts, \& Sciences, University of Southern California, Los Angeles, CA 90089, USA}
\affiliation{Center for Quantum Information Science \& Technology, University of
Southern California, Los Angeles, CA 90089, USA}
\author{D. M. Hartsell}
\affiliation{Department of Physics \& Astronomy, Dornsife College of Letters, Arts, \& Sciences, University of Southern California, Los Angeles, CA 90089, USA}
\affiliation{Center for Quantum Information Science \& Technology, University of
Southern California, Los Angeles, CA 90089, USA}
\author{E. Vlachos}
\affiliation{Department of Physics \& Astronomy, Dornsife College of Letters, Arts, \& Sciences, University of Southern California, Los Angeles, CA 90089, USA}
\affiliation{Center for Quantum Information Science \& Technology, University of
Southern California, Los Angeles, CA 90089, USA}
\author{H. Zhang}
\affiliation{Department of Electrical Engineering, Viterbi School of Engineering, University of Southern California, Los Angeles, CA 90089}
\affiliation{Center for Quantum Information Science \& Technology, University of
Southern California, Los Angeles, CA 90089, USA}
\author{E. M. Levenson-Falk}
\email[Corresponding author: ]{elevenso@usc.edu}
\affiliation{Department of Physics \& Astronomy, Dornsife College of Letters, Arts, \& Sciences, University of Southern California, Los Angeles, CA 90089, USA}
\affiliation{Center for Quantum Information Science \& Technology, University of
Southern California, Los Angeles, CA 90089, USA}

\date{\today}

\begin{abstract}
Nonequilibrium quasiparticles are ubiquitous in superconducting electronics. These quasiparticles can trap in the internal Andreev bound states of a phase-biased Josephson junction, providing a mechanism for studying their presence and behavior. We characterize a quasiparticle trapping detector device based on a two-junction aluminum nanobridge superconducting quantum interference device incorporated into a transmission-line resonator. When flux-biased, distinct resonant frequencies develop depending on the trapped quasiparticle number. We demonstrate continuous detection of up to 3 trapped quasiparticles, with detection of a trapped quasiparticle with signal-to-noise ratio of 27 in 5 $\mu$s. We describe initial measurements of quasiparticle behavior and discuss the possible optimization and application of such detector devices.
\end{abstract}

\maketitle

Superconducting qubits and other low-temperature superconducting electronics have ubiquitous populations of quasiparticles (QPs) far above their thermal equilibrium prevalence \cite{Aumentado2004,Martinis2009a,Catelani2011b,Rainis2012,Wang2014a,Vool2014,Glazman2021a}. These QPs can cause loss\cite{Gustavsson2016}, spurious excitation\cite{Serniak2018a}, and spectral noise when tunneling across qubit junctions. Even when QP populations are extraordinarily low \cite{Somoroff2021}, rare bursts of QPs can induce correlated errors that are difficult to address with error correction algorithms \cite{Vepsalainen2020a,McEwen2021}. QPs may be generated by stray infrared photons \cite{Barends2011}, cosmic rays and other high-energy radiation sources \cite{Vepsalainen2020a,Cardani2021,McEwen2021,Wilen2021}, or materials defects \cite{Kurter2021}. Many experiments have probed QP behavior via their tunneling across Josephson junctions in charge-sensitive transmons \cite{Sun2012,Riste2013a,Serniak2018a,Serniak2019a}, giving valuable insight into their effects on qubits. However, these measurements are discrete and cannot distinguish between 0 and 2 tunneling events. Trapping measurements \cite{Zgirski2011,Levenson-Falk2014a,Janvier2015,Hays2018,Hays2020} provide a tool for continuous, nonsaturating monitoring of QP behavior. Furthermore, QP traps have been proposed as a tool to mitigate QPs' effects on qubits \cite{Sun2012,Wang2014a,Riwar2016,Martinis2021}, as QPs may diffuse great distances after being generated \cite{Patel2018,McEwen2021,Wilen2021}. The trapping process itself is thus worthy of study, in addition to providing insight into bulk QP behavior.\par

In this letter we characterize a device optimized for continuous, non-saturating measurements of QP trapping in Andreev states. Using microwave reflectometry we are able to continuously detect 0, 1, and 2 or more trapped quasiparticles in 5 us with a signal-to-noise ratio (SNR) of 27. By altering the detector bias we are also able to distinguish 3, 2, and 1 or fewer QPs. We discuss straightforward improvements that can further improve SNR and allow detection of many more trapped QPs at a single bias point. Our device provides a prototype for detectors optimized for continuous measurements of QP behavior and properties and for studies of the dynamics of Andreev states coupled to resonant cavities.\par

In the semiconductor picture of the Josephson effect, supercurrent is carried by electrons/holes traveling in 1 dimensional conduction channels. At the junction boundaries, the electron (hole) reflects as a hole (electron) \cite{Andreev1964a}. This Andreev reflection causes a $\pm 2e$ charge transfer, where $e$ is the elementary charge, and thus transmits a Cooper pair across the junction. Each channel forms a pair of Andreev bound states with energies 
\begin{equation}
    E_{A\pm} = \pm\Delta \sqrt{1-\tau \sin^2 \frac{\delta}{2}}
\end{equation}
where $\Delta$ is the superconducting gap, $\tau$ is the transmittivity of the channel, and $\delta$ is the phase bias across the junction. 
In the semiconductor picture the Fermi energy is 0, so at temperatures $T \ll \Delta/k_B$, normally the upper Andreev state is unoccupied and the lower state is occupied, carrying the supercurrent across the junction. This is the channel's $|g\ket$ state with total energy $-E_A$. QPs in the bulk only exist at energies greater than $\Delta$ (or as unoccupied states below $-\Delta$), so it is energetically favorable for a QP to drop into the unoccupied upper Andreev state, bringing the channel to the $|o\ket$ state with 0 energy. This ``poisons'' the channel, eliminating it from carrying supercurrent and increasing the Josephson inductance. The lower-level QP may also be promoted to the upper Andreev level, bringing the channel to the $|e\ket$ and producing twice the inductance shift of the $|o\ket$ state; or this QP may be cleared from the junction completely, creating another degenerate 0-energy $|o\ket$ state with 0 supercurrent \cite{supplement}. Researchers have demonstrated continuous monitoring of QPs trapping in point contact \cite{Janvier2015} and semiconductor nanowire \cite{Hays2018,Hays2020} junctions; however, these junctions have few channels and so quickly saturate as QP detectors.

Three dimensional aluminum nanobridge Josephson junctions achieve good phase confinement and nonlinearity in an all-superconducting design \cite{Vijay2009,Vijay2010,Levenson-Falk2011}. Importantly, nanobridges comprise many conduction channels ($\sim 100-1000$) with $\tau \sim 1$, approximately following the Dorokhov distribution $\rho(\tau) = \frac{\pi \hbar G}{2 e^2}\frac{1}{\tau\sqrt{1-\tau}}$, where $G$ is the junction's normal-state conductance \cite{Dorokhov1982}. When 2 identical nanobridges are placed in a loop, forming a superconducting quantum interference device (SQUID), the junctions' phase bias $\delta$ is simply set by the flux bias $\Phi$ as $\delta = \pi \phi$, where $\phi\equiv \Phi / \Phi_0$ and $\Phi_0$ is the flux quantum. In an Al nanobridge SQUID near half-flux ($\delta \approx \pi/2$, a channel with $\tau \approx 1$ has $E_{A+}/h \approx 29$ GHz, or a trap depth of $(\Delta - E_A)/h \approx 12$ GHz, far greater than the thermal energy at 15 mK. These junctions thus function as QP traps with many deep trap states when they are phase-biased.\par

Our device consists of a co-planar waveguide (CPW) resonator in which the center trace is terminated by a nanobridge SQUID; see Fig.~\ref{fig:device}(a) and (b). Fabrication details are given in the supplementary material\cite{supplement}. The fundamental (quarter-wavelength) mode of our resonator with zero flux through the SQUID is at  $\omega_0(0) = 2\pi\times 4.302$ GHz and has linewidth $\kappa = 2\pi\times 250$ kHz, largely set by the coupling to the microwave feedline. This device was imaged and found to have junctions that appear nearly identical visually; past studies indicate they may thus be treated as symmetric\cite{Vijay2010,Levenson-Falk2011}. Trapping in either junction produces a similar resonant frequency shift, and so for QP detector applications they may be thought of as a single junction with twice the number of channels. We note that the resonator energy per photon is much smaller than the trap depth at high flux bias, so absorption of resonator photons should not appreciably affect the quasiparticle states \cite{supplement}.

\begin{figure}[ht!]
    \centering
    \includegraphics{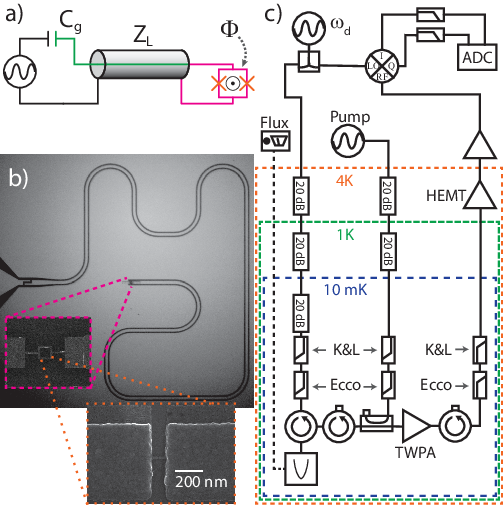}
    \caption{\footnotesize{[color online] (a) Device schematic. A CPW resonator (green) is grounded via a two-junction Al nanobridge SQUID (magenta). Flux bias through the SQUID phase-biases the junctions as $\delta = \pi \phi$. (b) Optical image of the device with inset SEM images of the SQUID (magenta) and a nanobridge junction (orange). (c) Simplified measurement schematic. A tone at $\omega_d$ continuously drives the flux-tunable resonator. The reflected signal is amplified by a TWPA at base stage followed by a HEMT at 4K and room temperature amplifiers. Isolators between HEMT and room temperature amplifiers are not shown to conserve space. The amplified signal is homodyne demodulated and I and Q components are low-pass filtered at 15 MHz, then digitized by an Alazar ATS9371 at 300 MSa/s and down-sampled to 10 MSa/s before saving. Microwave lines are optionally filtered at base stage by K\&L 12 GHz and custom Eccosorb 110 low-pass filters.}}
    \label{fig:device}
\end{figure}

Our measurement setup is shown schematically in Fig.~\ref{fig:device}(c). A signal generator provides a drive tone at $\omega_d(\phi) \approx \omega_0(\phi) - \kappa/2$. A power splitter sends half of this power into the dilution refrigerator where it is attenuated then (in some cooldowns) filtered by K\&L 12 GHz and Eccosorb low pass filters. The drive tone is circulated to reflect off our device, which is flux tunable via a DC coil in the packaging, and amplified by a travelling wave parametric amplifier\cite{Macklin2015} (TWPA) which is pumped at 8.078 GHz. The reflected signal is further amplified before IQ demodulation with reference to the original signal. The in-phase (I) and quadrature (Q) components are 15 MHz low pass filtered before digitization at 300 MSa/s. Raw data is down-sampled to 10 MSa/s before saving.\par

\begin{figure}[ht!]
    \centering
    \includegraphics{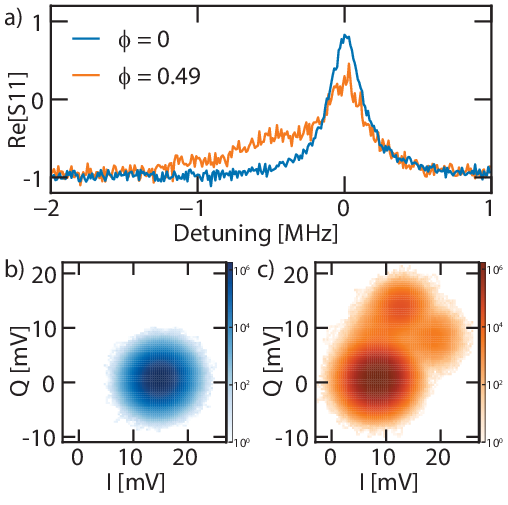}
    \caption{\footnotesize{[color online] (a) Ensemble measurement of resonator response at 0 flux (blue) and at $\phi=0.49$ (orange). When $\phi = 0.49$, distinct peaks are visible roughly 0.5 and 1 MHz below resonance, corresponding to 1 and 2 trapped QPs, respectively. (b) Histogram of continuous IQ data taken at 0 flux for 30 s. Data has 10 MHz sample rate and has been convolved with a Gaussian window with effective integration time of 3 $\mu$s. (c) Data taken at $\phi = 0.47$ with the same procedure as panel b. The darkest mode is due to the resonance with 0 trapped QPs. The second darkest, located near (I,Q) = (12 mV, 15 mV), corresponds to 1 trapped QP and the last mode corresponds to 2 or more trapped QPs (as this mode corresponds to the resonance moving far from the drive frequency).
    }}
    \label{fig:VNA}
\end{figure}

We first characterize our device with ensemble-averaged vector network analyzer (VNA) measurements of the resonance. Fig.~\ref{fig:VNA}(a) shows resonance measurements at flux biases of $\phi = 0$ and $\phi = 0.49$, taken on a cooldown in which the K\&L and Eccosorb filters were removed. The $\phi = 0.49$ trace in orange shows two shallow peaks at $\sim$ 0.5 and 1 MHz below the main resonance. These are the resonance peaks with 1 and 2 trapped QPs, respectively, showing the resonance shifting due to the change in nanobridge inductance. These ensemble measurements average over all possible QP trapping configurations, and so a resonance peak amplitude corresponds to the probability of that configuration. We then move on to time-domain IQ measurements as described above. Panels (b) and (c) are log-scale histograms showing 30 seconds of continuous IQ data for $\phi = 0$ (in blue) and $\phi = 0.47$ (in orange), respectively. The data shown has been integrated by convolving with a Gaussian window of effective integration time $\sqrt{2\pi\sigma} = 3 ~\mu$s. 
In the finite-flux data of panel (c) we can immediately see 3 distinct modes with excellent separation in the log scale plots. The darkest peak (lower left), with the most counts by far, is due to the response with 0 QPs in Andreev traps. The next darkest (upper center) is from having 1 QP trapped while the lightest mode (middle right) is from 2 QPs in Andreev traps and/or excitation of a single channel into the $|e\ket$ state. Trapping of more than 2 QPs moves the resonance multiple linewidths and thus saturates the change in response to additional trapping, so these counts lie on top of the 2-QP distribution. We later discuss methods for avoiding this saturation. By probing at a frequency close to the 2-QP resonant frequency, we are able to observe 0, 1, 2, and 3 QP modes, confirming that we do indeed see multiple QPs trapping and not simply the $|e\ket$ state of a single channel; data is shown in the supplementary material \cite{supplement}. We have verified that these modes are indeed due to Andreev trapping of QPs by measuring the weights of each mode in the presence of a ``clearing tone'' at 17 GHz \cite{supplement}. A trapped QP may absorb a photon from this tone and so be promoted back into the bulk continuum; the frequency was chosen to be greater than the trap depth and because it happens to couple efficiently into the device. We find that the tone causes the 1- and 2-QP mode counts decrease while the 0-QP mode counts increase. We have also observed that both the separations and the weights of the modes increase as a function of flux, which agrees qualitatively with a QP trapping picture \cite{supplement}.\par

\begin{figure*}[ht!]
    \centering
    \includegraphics{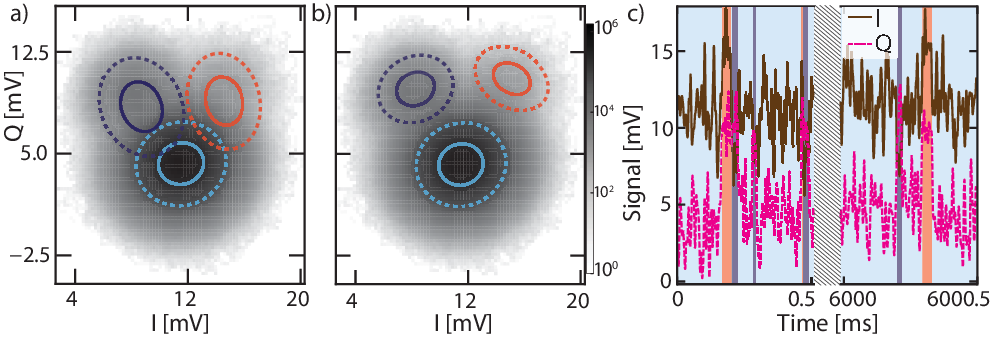}
    \caption{\footnotesize{[color online] (a) Initial clustering of 5 $\mu$s integrated data using the scikit-learn Gaussian mixture module produces modes with 1$\sigma$ (solid) and 2$\sigma$ (dashed) contours for 0, 1, and 2 or more trapped QPs in light blue, dark blue, and orange, respectively. (b) Subsets of the data in which the occupation is constant for a long time ($4\langle\tau_i\rangle$) are individually fit to Gaussian distributions. Means and covariances of each mode are then fixed and the full dataset is fit with mode weights as the only free parameters. (c) Two sections of time series data with I in brown and Q in dashed magenta. The background color is light blue, dark blue, and orange for 0, 1, and 2+ trapped QPs, respectively. All data taken at $\phi = 0.47$.}}
    \label{fig:detection}
\end{figure*}

We now turn to extracting the QP trap occupation as a function of time from the continuous IQ data. To optimize our analysis procedures we choose data that stresses the detector's capabilities. We use data from a cooldown with the K\&L and Eccosorb filters, which reduces QP generation by infrared radiation and thus shows a much lower QP number than that shown in Fig.~\ref{fig:VNA} \cite{Serniak2018a}, and use a low resonator drive power chosen to ensure the drive does not affect the QP configuration \cite{supplement}. We increase the integration time to 5 $\mu$s to compensate for this loss of SNR. We first fit histograms using a Gaussian mixture expectation-maximization algorithm implemented by the Python module available from scikit-learn\cite{scikit-learn}. This module takes in a subset of data, assigns each point to one of the specified number of modes, then tweaks assignments and mode parameters to maximize the total likelihood for all data and all modes. The result is a set of 3 Gaussian modes describing the data, shown by their 1-$\sigma$ (solid) and 2-$\sigma$ (dashed) contours overlaying the histogram in Fig.~\ref{fig:detection}(a). We then assign each time-series point to the mode with the highest posterior probability after Bayesian updating with a 3 sample rolling memory window to update the prior\cite{supplement}. Unfortunately, the Gaussian mixture fit is unreliable in terms of quality of fit and reproducibility of Gaussian mode parameters, which can vary significantly even when refitting the same data with the same initial guess. This unreliability is apparent from the fit distributions shown, which do not faithfully represent the means of the 1- and 2-QP modes.\par

To improve the fits we need to ``initialize'' the trapped QP configuration, thereby isolating each Gaussian mode for independent fitting. This is challenging, as we have no direct control of the trapped QP number. Fortunately, the Gaussian mixture procedure assigns most data points to the correct occupation. We use this initial assignment to fit the mean lifetime of each mode $\langle\tau_i\rangle$ and then identify periods in the time series when the extracted QP occupation is stationary for at least $4\langle\tau_i\rangle$. By stitching these ``quiet periods'' together, we build up large distributions of data points that are pre-assigned to modes. Each distribution is then independently fit to a Gaussian to fix its mean and covariance. Finally, the full dataset is fit to a mixture of 3 Gaussians with these same means and covariances, with the weight of each mode as the only free parameters. Fig.~\ref{fig:detection}(b) shows the result, with 1-$\sigma$ (solid) and 2-$\sigma$ (dashed) contours overlaying the data histogram. It is immediately evident that the ``quiet periods'' method produces a better quality of fit.  

We extract QP occupation from time series as before with updated Gaussian parameters. Fig.~\ref{fig:detection}(c) shows two 0.5 ms sections of data with a 5 $\mu$s Gaussian convolution. These sections were chosen to demonstrate switching events and are far more ``active'' than typical data, which mostly stays in the 0 QP mode. The background shading represents the extracted QP trap occupation (light blue for 0 QPs, dark blue for 1, and orange for 2 or more). Transitions between the 3 configurations are clearly visible and appear to be faithfully captured by the assignment algorithm. The detector's SNR at our operating parameters ($\sim$ 4 photon drive power, 5~$\mu$s integration), defined as the separation of mode centers in the IQ plane squared divided by product of their standard deviations along the line between them, is 27 for $0-1$ distinguishability, 32 for $1-2$, and 30 for $0-2$. The $0-1$ SNR gives a detector noise floor of $6.1 \times 10^{-4}$ QPs /$\sqrt{\text{Hz}}$, i.e., we can detect 0.00061 of the signal from a QP trapping with SNR = 1 after 0.5 s of integration, assuming a stationary occupation.

We now briefly describe the QP behavior measured with our device.  We see a mean trap occupation of $n_{qp} = 0.0185$, state occupation probabilities of $P_0 = 0.983$, $P_1 = 0.0155$, $P_2 = 0.00148$, and state lifetimes of $\tau_0 = 728~\mu\mathrm{s}$, $\tau_1 = 12.7~\mu\mathrm{s}$, $\tau_2 = 4.73~\mu\mathrm{s}$. These lifetimes are corrected for the detector bandwidth assuming Poisson switching processes \cite{Aumentado2006,supplement}. The distributions of lifetimes do appear Poissonian in the long-time limit, but better SNR (discussed below) may resolve fast events which may show non-Poisson behavior. We also note that we see transitions between all three of the 0, 1, and 2 QP modes. We attribute the $0-2$ transitions to either correlated trapping of 2 QPs in less than a detector bandwidth or to direct $|g\ket \rightarrow |e\ket$ excitation of a single channel; spectroscopic measurements of the trapped QPs should be able to distinguish between these processes \cite{supplement}. Future work will probe switching rates as a function of bias and environmental parameters (e.~g.~flux and temperature), analyze correlations between switching events, and develop more sophisticated state-assignment algorithms that do not assume independent (Poisson) switching.

We note that our device is not fully optimized for high sensitivity. While the resonant frequency was kept low in order to be less than the trap depth, raising the resonance slightly to $\sim$ 6 GHz by shortening the waveguide would increase the participation ratio of the nanobridge inductance to total inductance while still remaining far below trap-clearing frequencies. Similarly, reducing the resonator's characteristic impedance from 50 $\Omega$ to an easily-achievable $\sim 30~\Omega$ would further reduce the linear inductance, increasing sensitivity. Additionally, the TWPA used as a first-stage amplifier had a moderate $\approx$ 15 dB gain (due to being operated near the edge of its bandwidth) and adds noise above the quantum limit. Adding a standard parametric amplifier with a near-quantum-limited noise temperature and 20 dB of gain as a preamplifier will further improve SNR. We may also trade off some of this sensitivity and increase the resonator bandwidth, thus allowing for detection of more than 2 QPs without saturating the response. We also note the possibility of detecting higher numbers of trapped QPs by probing the device with multiple probe tones simultaneously. By performing heterodyne measurement of probe tones centered on, e.g., the 1-, 3-, and 5-QP resonant frequencies, and correlating the measured outcomes, we should be able to detect up to 6 trapped QPs with a similar device.\par
In conclusion, we have developed a device for the ultra-low-noise continuous detection of up to 2 quasiparticles trapping in Andreev bound states. Our device is capable of detecting a trapped QP with SNR of 27 in 5 $\mu$s, giving it a noise floor of $6.1 \times 10^{-4}$ QPs /$\sqrt{\text{Hz}}$. Straightforward extensions are possible to higher sensitivity and QP saturation number. Our device can be used for QP studies including statistical analysis of trapping and untrapping rates and trap occupation, spectroscopic measurements of trapped QP energy distributions, effects of environmental variables such as temperature, and testing of QP mitigation techniques.\par
The authors would like to acknowledge J. Aumentado and L. Glazman for useful discussions. We acknowledge MIT Lincoln Laboratory and IARPA for providing the TWPA used in this work. This work was funded by the AFOSR YIP under grant FA9550-19-1-0060 and by the NSF DMR under grant DMR-1900135. All data presented is available upon request to the corresponding author.


\bibliography{QuasiparticlesAPL}

\pagebreak
\clearpage
\clearpage
\setcounter{figure}{0}
\setcounter{equation}{0}
\setcounter{page}{1}
\renewcommand{\thefigure}{S\arabic{figure}}
\renewcommand{\theequation}{S\arabic{equation}}
\widetext

\section{Supplement to ``Continuous Real-Time Detection of Quasiparticle Trapping in Aluminum Nanobridge Josephson Junctions''}

\subsection{Fabrication details}
We start with a high-resistivity intrinsic silicon wafer with 5/50~nm Ti/Au markers bordering a 1~cm square chip. AZ1512 photoresist is spun on at 4000~RPM for 60 seconds, followed by a pre-exposure bake at 100$^\circ$C for 60 seconds. We use shadow-mask photolithography to pattern the coplanar waveguide resonator and the chip is developed in AZ400K at room temperature for 45 seconds. Aluminum is deposited via e-beam evaporation to a thickness of 100~nm. Liftoff is performed by soaking in acetone for 30 minutes followed by a 30 second sonication.

To fabricate the junctions, we next spin the chip with MMA EL6 at 3200~RPM for 60 seconds followed by a 5 minute bake at 170$^\circ$C. This results in a layer of MMA around 100~nm thick. We then spin a layer of PMMA A2 at 3600~RPM followed by a 5 minute bake at 180$^\circ$C to achieve a 60~nm layer of PMMA on top. The SQUID is patterned with e-beam lithography using a beam current of $\sim$18~pA and a 2~nm step size, with some intentionally low dose elements on one side of the pattern that define undercuts – regions where the MMA develops away but a lip of PMMA remains. The junctions are defined as 100~nm long lines bridging the two sides of the SQUID. The chip is developed in a -15$^\circ$C chilled beaker of MIBK:IPA 1:3 for 60 seconds, followed immediately by drying under a strong jet of N$_2$. The timing is crucial as any delay between development and drying will warm the developer and result in overdeveloped features. Aluminum is deposited via e-beam evaporation to 8~nm at direct incidence, then the substrate stage is tilted to 35 degrees and another 62~nm thickness is accumulated on chip. This angled evaporation is necessary to define the 3 dimensional nanobridge as the 35 degree incidence blocks the line of sight to the substrate in the junction line while allowing it in the SQUID leads. Liftoff is performed by a 2 hour soak in acetone warmed to 45$^\circ$C, followed by a 10 second sonication.

The last step is to make superconducting contact between the resonator and the SQUID. AZ1512 is spun on as before and photolithography defines a pair of rectangles for each device. This is developed in AZ400K as before. Each pair of rectangles opens a window that covers a portion of the SQUID contact pad and the resonator. The chip is then loaded into the e-beam evaporator and an end-Hall ion source running Argon is used to remove surface oxides from the SQUID and resonator portions exposed in these windows. Our recipe is 80~V at 3.5~A discharge (resulting in $\sim$50~eV neutral Argon bombardment) at 35 degree incidence for 270 seconds. We find that this is enough to remove the native oxide layer while not etching through the thinnest (70~nm) layer of Aluminum. Without breaking vacuum, we deposit 250~nm of Al to ensure good contact between resonator and SQUID. Liftoff is performed by a 30 minute soak in acetone followed by 10 second sonication.

\subsection{Fitting flux tuning and participation ratio}

To characterize the junction properties we fit the device resonant frequency, $\omega_0$, as a function of applied flux, $\phi$, and extract the Josephson inductance participation ratio at zero flux, $q_0$. To fit this, we need to derive the flux tuning equation. We assume we have two symmetric junctions which each approximately obey the KO-1 current-phase relation (CPR) of an ideal short diffusive metallic weak link~\cite{KO1},

\begin{equation}
    I(\delta) = \frac{\Delta N_e}{4 \varphi_0} \cos\frac{\delta}{2} \tanh^{-1}(\sin \frac{\delta}{2}),
\end{equation}

where $\Delta$ is the superconducting gap, $N_e$ is the effective number of channels, $\varphi_0$ is the reduced flux quantum, and $\delta = \pi \phi$ is the phase bias across the junction. We note that this approximation of symmetric junctions has been shown to be valid for devices with nanobridges which appear visually similar to each other, as our device's nanobridges do \cite{Vijay2010,Levenson-Falk2011}. This gives a SQUID inductance,

\begin{equation}
    L_S(\phi) = \frac{L_{j,0}}{2}[1-\sin\frac{\pi \phi}{2}\tanh^{-1}(\sin\frac{\pi \phi}{2})],
\end{equation}

where the Josephson inductance per bridge at zero flux is

\begin{equation}
    L_{j,0} \equiv \frac{8 \varphi_0 ^2}{\Delta N_e}.
\end{equation}

With the definition of the Josephson inductance participation ratio as 
\begin{equation}
    q(\phi) = \frac{L_S(\phi)}{L+L_S(\phi)}
\end{equation}

and $q_0 \equiv q(0)$, we can rewrite the resonant frequency function as 

\begin{equation}
    \omega(\phi) = \omega_0 \left[1 + q_0 \frac{\sin\frac{\pi \phi}{2} \tanh^{-1}(\sin\frac{\pi \phi}{2})}{1-\sin\frac{\pi \phi}{2} \tanh^{-1}(\sin\frac{\pi \phi}{2})} \right]^{-1/2}.
\end{equation}

\begin{figure}[ht!]
    \centering
    \includegraphics[width = 0.75\textwidth]{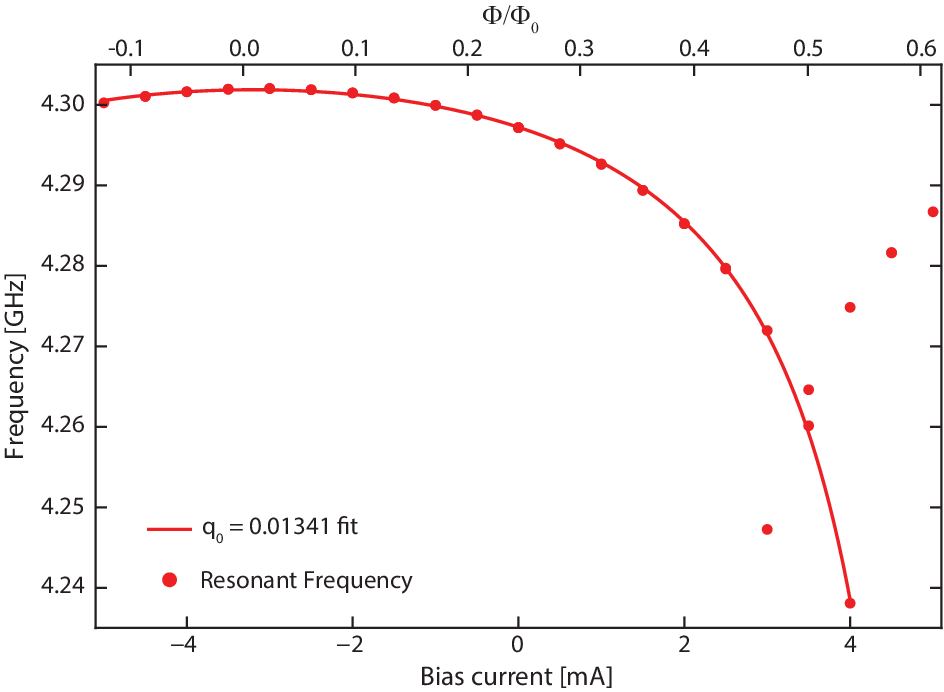}
    \caption{The extracted resonant frequency as a function of flux. The solid curve is the fit to the equation in the supplemental text. We extract a Josephson inductance participation ratio at zero flux of 0.01341. We see hysteretic branches in the flux tuning, as expected for a SQUID with junctions which have non-sinusoidal CPRs.}
    \label{part_ratio}
\end{figure}

We measure $S_{11}$ as a function of applied flux to extract the resonant frequency, then fit it to the above equation. The results for the device in the main text are presented in Fig.~\ref{part_ratio}. We find our Josephson participation ratio at zero flux is $q_0 \approx 0.01341$. We obtain the resonator’s linear inductance of $L = 1.83$ nH via finite element EM simulations. From this and the participation ratio we extract the SQUID inductance at zero flux as $L_S(0) \approx 24$~pH, corresponding to an effective channel count $N_e \approx 660$ per junction.

We note that our 100~nm long nanobridges have CPRs which are not quite equal to the KO-1 formula~\cite{Vijay2009, Vijay2010, Levenson-Falk2011}. However, we have found these junctions’ CPRs are very well approximated by an ideal KO-1 junction in series with a small linear inductance. This approximation leads to an insignificant modification of the participation ratio, as this small extra linear inductance simply adds to the large resonator linear inductance, and so we neglect it. We also note that the flux tuning is hysteretic in a small range around $\phi = 0.5$. This is expected for any SQUID incorporating junctions with a non-sinusoidal CPR, and is well fit by our model \cite{Levenson-Falk2011}. We conduct most measurements at a flux of $\phi = 0.47$ to ensure deep traps while remaining in the favored branch.

\subsection{Power dependence}

\begin{figure}[ht!]
    \centering
    \includegraphics[width = 0.75\textwidth]{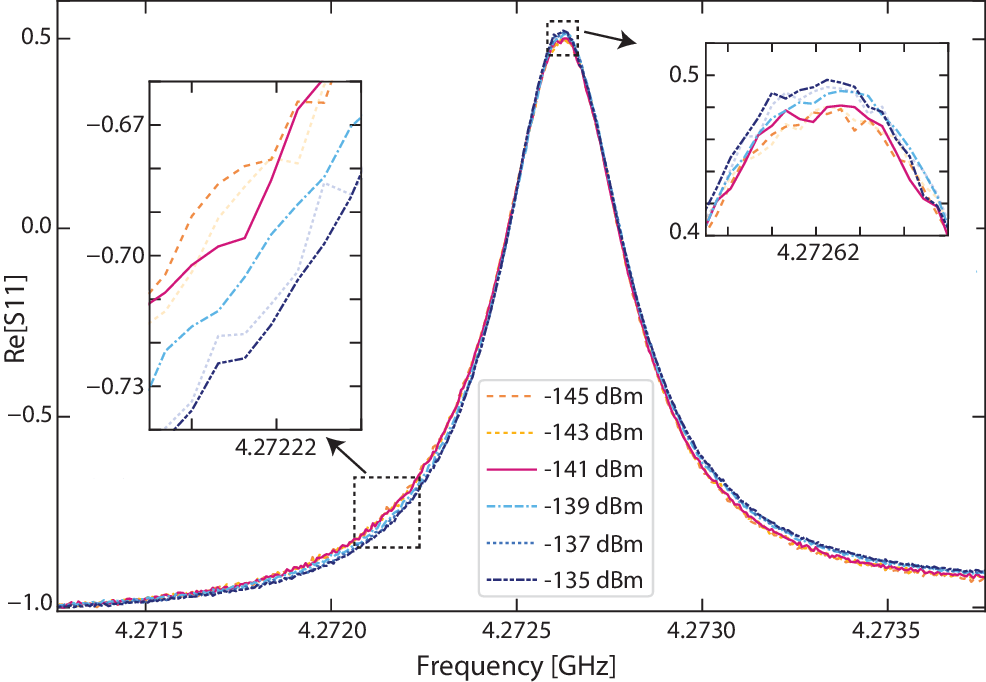}
    \caption{Ensemble-averaged resonator response at $\phi=0.47$ as a function of measurement power. At powers below -141 dBm the response is roughly power-independent. As we increase the measurement power beyond -139 dBm, the weight of the 1-QP mode begins to decrease and the weight of the 0-QP mode increases.}
    \label{power_dependence}
\end{figure}

The approximately 4.3 GHz resonant frequency of our device means no single resonator photon has enough energy to excite a QP out of a deep trap state. However, multi-photon processes can occur, as there is a finite probability of multiple resonator photons for any strength of coherent drive. To find a measurement tone power that does not affect QP trapping, we take ensemble-averaged VNA measurements of the resonance as a function of probe power. Results at $\phi = 0.47$ are shown in Fig.~\ref{power_dependence}. We find that at measurement powers below -141 dBm the resonator response is power-independent. As we increase power above this level, we find that the weight of the response near the 1-QP resonant frequency begins to decrease and the weight near the 0-QP resonant frequency begins to increase. We attribute these changes to clearing of QPs from the trap states. To optimize SNR while ensuring the detector is not affecting the QP configuration, we choose a measurement power just below -141 dBm. We note that at higher measurement powers the resonator response becomes nonlinear, leading to increased sensitivity per QP at the expense of bandwidth and saturation at a lower QP number. However, in our device this nonlinear regime happens at powers which are already affecting the QP configuration, and so we neglect it in our analysis.

\subsection{Quiet periods Gaussian fit and occupation extraction}

We first fit the IQ voltage-vs-time data using the scikit-learn Gaussian Mixture module described in the main text to find the conditional probability of a measurement outcome given a mode $P(\vec{x}|M_i)$ and the prior probability of each mode $P(M_i)$. We then run this through a Bayesian state identification algorithm (described below) to extract a rough occupation. The next task is to separate out subsets of data in which the mode (i.e.~the trapped QP number) is stationary for a significant portion of time. To do this, we first fit the lifetimes of each occupation by building a histogram of the times spent in each mode. We take the portion of this histogram above the mean and fit it to an exponential to extract the apparent lifetime of the mode $\tau_i^*$. For each mode, we separate out only those subsets of data in which the mode is constant for longer than $4\tau_i^*$. We trim the first and last 5\% of the lifetime from each subset before stitching them together, which should exclude any ringdown effects. This provides 3 datasets, one for each mode, where we can confidently assign the data to a certain QP occupation.

We separately histogram each data set and fit each to a 2D Gaussian distribution. Using the means and covariances of each fit as fixed values, we perform another fit of a histogram of the original full dataset to a sum of all 3 Gaussians. In this fit, the relative weights of the 3 modes are the only free parameters. This final step is of critical importance as the weights the modes ($P(M_i)$) are a significant contributor to the calculation of posterior probabilities used in extracting the occupation. 
Since we expect the resonator to remain in the same mode for many consecutive points in the time series, we apply a Bayesian inference scheme with a memory window of 3 points. i.e., the probability that the $k_{th}$ data point $\vec{x}_k$ belongs to the $i_{th}$ mode is
\begin{equation*}
P(M_i|\vec{x}_k,\vec{x}_{k-1},\vec{x}_{k-2}) = \frac{P(\vec{x}_k|M_i)P(M_i|\vec{x}_{k-1},\vec{x}_{k-2})}{\sum_{j=0}^2 P(\vec{x}_k|M_j)P(M_j|\vec{x}_{k-1},\vec{x}_{k-2})}
\end{equation*}
where 
\begin{equation*}
P(M_i|\vec{x}_{k-1},\vec{x}_{k-2}) = \frac{P(\vec{x}_{k-1}|M_i)P(M_i|\vec{x}_{k-2})}{\sum_{j=0}^2 P(\vec{x}_{k-1}|M_j)P(M_j|\vec{x}_{k-2})}
\end{equation*}
and $P(M_i|\vec{x}_{k-2})$ is the posterior probability of the Gaussian mode $M_i$ conditioned on the data, defined as
\begin{equation}
    P(M_i|\vec{x}_k) = \frac{P(\vec{x}_k|M_i)P(M_i)}{\sum_{j=0}^2 P(\vec{x}_k|M_j)P(M_j)},
\end{equation}
Finally, the quasiparticle occupation at the $k_{th}$ point in the time series is taken as the mode $M_i$ with maximum $P(M_i|\vec{x}_k,\vec{x}_{k-1},\vec{x}_{k-2}) \; \forall \; i \in \{0,1,2\}.$

\subsection{Clearing tone}

\begin{figure}[ht!]
    \centering
    \includegraphics[width=0.75\textwidth]{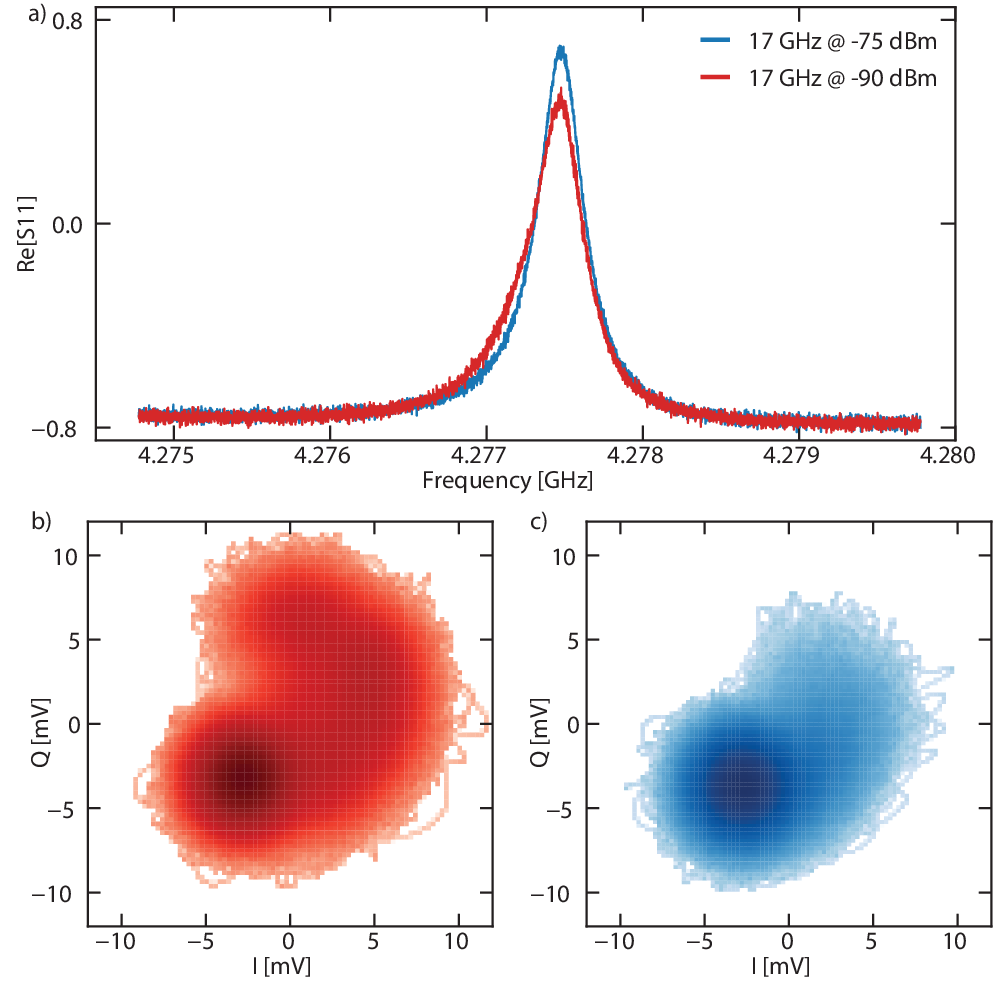}
    \caption{Microwave response of the resonance at flux bias 0.45 as a function of clearing tone power. The legend in (a) shows the approximate power of the clearing tone at the plane of the device. The red curve is at low power and shows a significant bump on the low frequency side due to averaging over many configurations of QP occupation. At high clearing tone power (blue curve) we see the resonator response narrows, becomes more symmetric, and appears to be more over-coupled. We stress that the ratio of internal loss to external coupling is not actually changing; rather, the response is taller because we are no longer averaging over other configurations of QP occupation. (b) Histogram of time series data with the clearing tone at the same power as the red trace in (a). (c) Histogram of time series data with the clearing tone at the same power as the blue trace in (a). Note that any trapped quasiparticles are quickly excited back above the gap, greatly reducing the weights of the 1- and 2-QP modes and increasing the weight of the 0-QP mode.}
    \label{clearing_tone}
\end{figure}

As described in the main text, we expect the trap depth to be $\Delta - E_A(\phi = 0.47) \approx10.7$~GHz. To test that our observed behavior is due to quasiparticles trapping in Andreev states, we attempt to clear these trapped quasiparticles by injecting a tone with sufficient energy to excite the trapped quasiparticles into the continuum of available states above the gap. Fig.~\ref{clearing_tone} shows the microwave response of the resonator as a function of a 17~GHz tone drive power. This frequency was chosen as it is higher than the trap depth and, due to details of our device and measurement setup, couples energy efficiently into the resonator. Ensemble measurements of the reflection coefficient show that, when driven strongly, the resonance curve recovers the single Lorentzian shape that it has at 0 flux. We also see that the second and third modes apparent in the continuous IQ measurements have greatly reduced amplitudes, indicating that QPs spend far less time in trap states as the 17~GHz tone clears them out. Combined with the qualitative flux dependence of trapping\textemdash trapping modes move farther apart and become stronger as the flux bias becomes deeper\textemdash we take this as definitive evidence of QP trapping, similar to that shown in ensemble measurements of similar devices~\cite{Levenson-Falk2014a}. We note that the frequency of the clearing tone can be swept to spectroscopically measure the exact energies of trapped QPs. We plan such spectroscopic measurements in future studies.

\subsection{Quasiparticle transition rates}
\begin{figure}[ht!]
    \centering
    \includegraphics[width=0.95\textwidth]{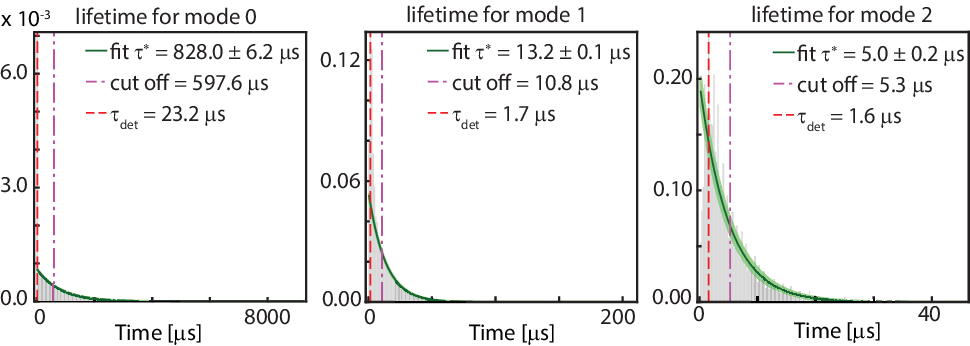}
    \caption{Histograms of mode lifetimes for the 0- (a), 1- (b), and 2-QP (c) modes. All lifetimes are exponentially distributed at long times, but show a distinct peak at low times which we attribute to the finite detection bandwidth $\tau_{det}$. We extract the apparent lifetimes $\tau_i^*$ by fitting the behavior above a cutoff time, defined as the mean measured lifetime. These lifetimes are then adjusted to correct for the finite detector bandwidth as described in the text, giving $\tau_0 = 728~\mu\mathrm{s}, \tau_1 = 12.7~\mu\mathrm{s}, \tau_2 = 4.73~\mu\mathrm{s}$.}
    \label{lifetimes}
\end{figure}

Using our extracted QP state as a function of time, we are able to determine the lifetimes of the transition rates. We show data in Fig.~\ref{lifetimes}. The distributions of lifetimes for each of the 0-, 1-, and 2-QP modes show an exponential dependence at long times, but show a distinct peak at low times. We attribute this peak to the finite detection bandwidth $\tau_{det,i}$ for detecting a switch out of the $i$th mode, and correct for it by assuming a Poisson process and following the method developed by Aumentado and Naaman\cite{Aumentado2006}. We first find the mean lifetime of all detector counts for a mode and set this as a cutoff time, fitting the distribution of lifetimes longer than this cutoff to an exponential. This gives an apparent state lifetime $\tau_i^*$ for the $i$th mode, which is generally longer than the true lifetime. We also extract the apparent lifetimes outside the $ith$ mode $\tau_{i,out}^*$ and the detection time for switching back to the $i$th mode $\tau_{det,out}$ We then adjust the mode lifetime using the formula
\begin{equation}
    u = \frac{1 - u^{*2} - v^{*2}}{1-u^*-v^*} u^* - u^{*2}
\end{equation}
Where $u = \tau_{det} / \tau_i$, $v^* = \tau_{det,out} / \tau_{i,out}^*$, $u^* = \tau_{det} / \tau_i^*$. This procedure gives the lifetimes quoted in the main text, $\tau_0 = 728~\mu\mathrm{s}$, $\tau_1 = 12.7~\mu\mathrm{s}$, $\tau_2 = 4.73~\mu\mathrm{s}$. We note that the finite detection SNR produces occasional ``blips'' from the 1 and 2 modes to the 0 mode and back, leading to a large population of short 0-mode lifetimes and likely biasing our results towards shorter 1- and 2-mode lifetimes. We are working to develop state assignment algorithms that are less sensitive to such events. Future work will also develop detector bandwidth corrections that do not rely on the Poisson assumption.

\subsection{Trapping and Excitation Mechanisms}
\begin{figure}[ht!]
    \centering
    \includegraphics[width=0.75\textwidth]{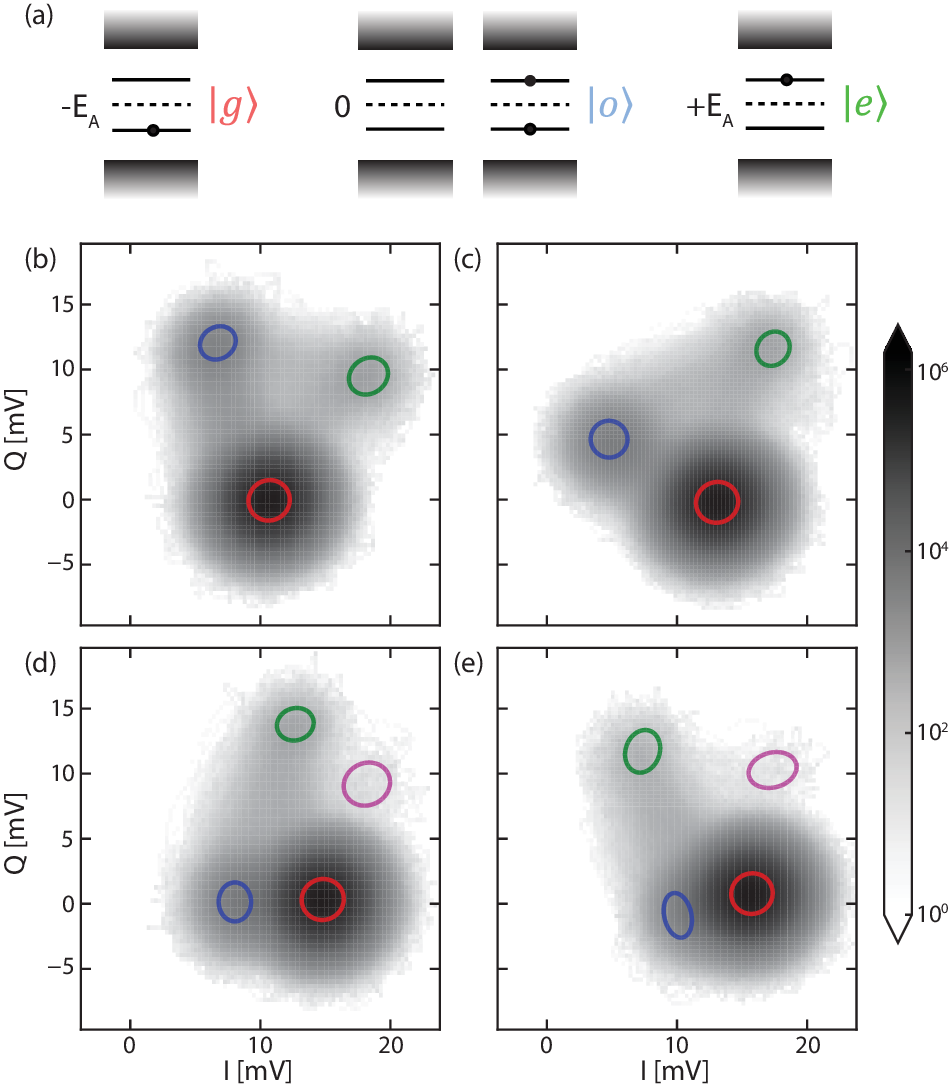}
    \caption{(a) Schematic description of the ground state $\ket{g}$, 2 degenerate first excited states $\ket{o}$, and second excited state $\ket{e}$ of a channel. Both $\ket{o}$ states produce the same resonant frequency shift, while the $\ket{e}$ state produces the same shift as 2 channels entering their $\ket{o}$ states. (b-e) Response of the device to different probe tone frequencies. The probe tone is stepped to progressively lower frequencies, starting with the midpoint of the 0- (red) and 1-QP (blue) mode resonances in (b) and ending with the 2-QP mode (green) frequency in (e). As the probe moves close enough, a 3-QP mode (purple) becomes apparent.}
    \label{higher_n}
\end{figure}

As discussed in the main text, a conduction channel can take on 3 states, shown schematically in Fig.~\ref{higher_n}(a). The ground state $\ket{g}$ with energy $-E_A$ has a QP in the lower Andreev state and an empty upper state. The 2 degenerate $\ket{o}$ states with 0 energy either have the lower QP excited out of its state or have an extra QP trapped in the higher state. The excited state $\ket{e}$ with energy $+E_A$ has the lower QP excited into the upper state. We note that both $\ket{o}$ states cause the same resonant frequency shift in our device, while a channel being excited to the $\ket{e}$ state has the same effect as two channels excited to their $\ket{o}$ states. As the gaps between Andreev states and between the lower Andreev state and the continuum are much larger than the gap between the upper state and the continuum, we anticipate that trapping of QPs in the upper states will be much more frequent than excitation from the lower state. However, we do see transitions between all 3 modes in our data. The 0-2 mode transitions are either the result of 2 QPs trapping in different channels (2 channels moving to their $\ket{o}$ states) in less than a detector bandwidth, or a single channel being excited to its $\ket{e}$ state. Clearing-tone measurements should be able to distinguish these processes: a clearing tone will take a single channel in $\ket{e}$ to $\ket{o}$, causing a frequency shift equivalent to clearing 1 QP. If 2 QPs are trapped, bringing 2 channels to $\ket{o}$, then a clearing tone will clear both and bring both channels back to $\ket{g}$, producing twice the frequency shift. We note that the presence of a 3-QP mode, discussed below, indicates that at least 2 and likely 3 channels (and QPs) are involved. We expect higher QP numbers to be involved in an environment with a higher background QP population, e.~g.~ one in which there is much more thermal photon radiation.

\subsection{Higher Number Detection}

In order to confirm that we really see multiple QPs trapping and not just a single channel excited to the $\ket{e}$ state, we tune our device to measure the 3-QP mode. Time series data in the main text is shown with the probe tone on or near the 1-QP resonant frequency. As the frequency shift per trapped QP is larger than the resonance linewidth, trapping 2 QPs fully saturates the response of the detector and the modes representing higher trap numbers will lie on top of the 2-QP mode in the IQ plane. However, we can rectify this issue straightforwardly by decreasing the frequency of the probe tone so that it lies closer to the resonances of these higher QP modes. This allows, for instance, distinguishing 2 vs 3 QPs at the expense of our ability to distinguish 0 vs 1. We show this procedure in Fig.~\ref{higher_n}. We step the probe frequency from halfway between the 0- and 1-QP resonant frequencies (panel b), to the 1-QP resonance (panel c), to halfway between 1- and 2-QP (panel d), to the 2-QP resonance (panel e). We initially see 3 modes, which we attribute to 0, 1, and 2 trapped QPs, circled in red, blue, and green respectively. As we move the probe frequency lower, a fourth mode (circled in purple) appears, breaking away from the previously-saturated 2-QP mode. We note that this data was taken with a higher measurement power which may affect the QP configuration and which caused the resonance to be slightly nonlinear. We used this higher power because the 3-QP mode was extremely rare in this cooldown (due to the low background QP density) and so was difficult see easily in a histogram if the SNR was not very high. This will not be a concern in situations where the mean trapped QP number is higher, or if the SNR improvements discussed in the main text are made.

\end{document}